\documentclass[12pt]{article}
\usepackage{amsmath,amsfonts,amssymb}

\def\ttau{\tilde{\tau}}

\def\tn{\tilde{n}}

\def\bA{\mathbf{A}}

\def\tlambda{\tilde{\lambda}}

\def\tx{\tilde{x}}

\def\be{\begin{equation}}

\def\hi{\hat{i}}
\def\hj{\hat{j}}
\def\ee{\end{equation}}

\def\bea{\begin{eqnarray}}

\def\eea{\end{eqnarray}}

\def\tl{\tilde{l}}

\def\tmH{\tilde{\mH}}

\def\mH{\mathcal{H}}

\def\tF{\tilde{F}}
\def\tA{\tilde{A}}

\def\tp{\tilde{p}}

\def\tA{\tilde{A}}

\newcommand{\tN}{\tilde{N}}

\def\tpi{\tilde{\pi}}

\newcommand{\mK}{\mathcal{K}}

\def\halpha{\hat{\alpha}}

\newcommand{\hk}{\hat{k}}

\def \bAi{\left(\mathbf{A}^{-1}\right)}

\def \bA{\mathbf{A}}
\def\ta{\tilde{a}}

\def\mH{\mathcal{H}}

\newcommand{\mL}{\mathcal{L}}

\def\pb #1{\left\{#1\right\}}

\begin{document}

	\begin{titlepage}

		\vskip 0.4 cm
		
		\begin{center}
			{\Large{ \bf Note About  D-Branes in Carrollian Background
			}}
			
			\vspace{1em}  
			
			\vspace{1em} J. Kluso\v{n} 			
			\footnote{Email addresses:
				klu@physics.muni.cz (J.
				Kluso\v{n}) }\\
			\vspace{1em}
			%			\textit{Department of Physics, University of Helsinki,
			%				P.O. Box 64,\\ FI-00014 Helsinki, Finland}\\
			%			\vspace{.3em} $^b$\textit{Department of Physics, Tafresh University, Tafresh, Iran}\\
			%			\vspace{.3em} 
			\textit{Department of Theoretical Physics and
				Astrophysics, Faculty of Science,\\
				Masaryk University, Kotl\'a\v{r}sk\'a 2, 611 37, Brno, Czech Republic}
			
			\vskip 0.8cm
			
			%			
			%			
			%			Josef Kluso\v{n}$\,^1$
			%			\footnote{Email address:
			%				klu@physics.muni.cz}\\
			%			\vspace{1em} $^1$\textit{Department of Theoretical Physics and
			%				Astrophysics, Faculty
			%				of Science,\\
			%				Masaryk University, Kotl\'a\v{r}sk\'a 2, 611 37, Brno, Czech Republic}\\
			%			
			%			
			%{\large Josef Kluso\v{n}$^{}$\footnote{E-mail: {\tt
			%klu@physics.muni.cz}} }
			%
			%\vskip 0.8cm
			%
			%{\it Department of
			%Theoretical Physics and Astrophysics\\
			%Faculty of Science, Masaryk University\\
			%Kotl\'{a}\v{r}sk\'{a} 2, 611 37, Brno\\
			%Czech Republic\\
			%[10mm]}
			
			\vskip 0.8cm
			
		\end{center}

		\begin{abstract}
This paper is devoted to the construction of stable and unstable Dp-branes
in generalized Carroll space-times. We also study tachyon solutions on the
world-volume of unstable Dp-brane in Carrollian background and discuss their physical interpretations.

		\end{abstract}
		
		\bigskip
		
	\end{titlepage}
	
	\newpage

%%%%%%%%%%%%%%%%%%%%%
%%%%Introduction %%%%%%%%%
%%%%%%%%%%%%%%%%%%%%
\section{Introduction and Summary}\label{first}
Recently the analysis of   small speed of light expansion of general relativity was performed in very interesting paper \cite{Hansen:2021fxi}. The main importance of this 
work is that this expansion was performed with the help of the perspective of non-Lorentzian geometry that allows elegant formulation of Carroll expansion of general relativity (GR). This formulation has its inspiration in the expansion of GR for large speed of light that leads to geometrical formulation of non-relativistic expansion of GR that was performed in \cite{VandenBleeken:2017rij,Hansen:2019pkl,Hansen:2020pqs}. Then systematic speed of light expansion was performed in \cite{Hansen:2021fxi} along these lines when small speed of light expansion leads to Carroll geometry with higher order geometrical fields. At the leading order we have well known duality between Newton-Cartan geometry and Carroll geometry
\cite{Duval:2014uoa,Hartong:2015xda}.

Another interesting question is to study  strings and p-branes in this Carrollian background. In fact, Carrollian strings and particles were previously analysed in
\cite{Bergshoeff:2015wma,Cardona:2016ytk,Kluson:2017fam}, see also  \cite{Bergshoeff:2020xhv,Roychowdhury:2019aoi}. These objects were defined by specific limiting procedure of the world-volume (world-sheet) fields that label positions of these objects in the target space-time together with rescaling of p-brane(string) tensions. It was shown in \cite{Bergshoeff:2015wma} that the dynamics of Carroll particle is trivial which was also confirmed in case of the string in \cite{Cardona:2016ytk}. Further, in \cite{Cardona:2016ytk}  $p-$brane
 Carrollian limit was  defined as rescaling $p+1$ longitudinal target space coordinates $x^\mu=cX^\mu$ and taking limit $c\rightarrow 0$. In this way an action for Carroll p-brane was found and it was also shown that its dynamics is trivial. This analysis was further generalized to the case of Carrollian limit of non-BPS Dp-brane that was performed in  \cite{Kluson:2017fam}.

The goal of this paper is to apply formalism developed in \cite{Hansen:2021fxi} to the case of Carrollian k-brane limit and find corresponding action for Dp-brane action in this background. Interestingly, we show that it is not necessary to have correlation between the number $k$ and $p$ which is in sharp contrast with the non-relativistic p-brane limit 
\cite{Kluson:2017abm,Blair:2021waq,Kluson:2020rij,Kluson:2019uza} where in order to cancel divergence that arises through non-relativistic limit in the kinetic term of p-brane we should have specific form of the $p+1$-gauge form that couples to world-volume of p-brane.  In case of Carrollian  $k-$brane limit we  find corresponding Dp-brane action in canonical formalism and we discuss its form 
in dependence on its world-volume dimensions and number of dimensions where Carrollian limit is performed. 

As the next step we extend this analysis to the case of non-BPS Dp-branes \cite{Sen:1999md,Bergshoeff:2000dq,Kluson:2000iy} that includes tachyon field on its world-volume together with the gauge field. We find the form of non-BPS Dp-brane action in Carrollian k-brane limit of gravity and study their equations of motion. We firstly consider tachyon kink solution following very nice analysis performed in \cite{Sen:2003tm} and further extended in 
\cite{Kluson:2005fj,Kluson:2005hd}. In case of non-BPS Dp-brane in Carroll background the situation is slightly different since the action is written in canonical form but we again show that the tachyon kink solution corresponds to the stable lower dimensional D(p-1)-brane in $k-$brane Carrollian background which is nice consistency check of the tachyon condensation. Another interesting situation occurs when we consider tachyon at its vacuum state. It was argued
in case of relativistic non-BPS Dp-brane in \cite{Sen:2003bc,Kim:2003he,Gibbons:2000hf,Kluson:2016tgp,Kwon:2003qn} 
that this vacuum solution with non-zero electric flux corresponds to the gass of relativistic strings. The same situation occurs in case of the Carroll non-BPS Dp-brane in flat space-time as was shown in  \cite{Kluson:2017fam}. We generalize this analysis to the case of non-BPS Dp-brane in  $k-$brane Carrollian background  and we show that the tachyon vacuum solution describes dynamics of fundamental string  in the same background.

Let us outline our results. We introduced $k-$brane Carrollian limiting procedure as generalization of the small speed of light expansion of general relativity that was performed in  \cite{Hansen:2021fxi}. Then we studied p-branes in this background and we showed that can be defined for any values of $k$ and $p$. We determined corresponding actions and discuss conditions under which they can be written in Lagrangian form. Then we studied    unstable Dp-branes 
in $k-$ brane Carrollian background. We determined corresponding equations of motion. We argued that the tachyon kink solution describes D(p-1)-brane in $k-$brane Carrollian background. Finally we also analysed tachyon vacuum solution and we argued that again corresponds to the gas of fundamental strings in $k-$brane Carrollian background.

This paper is organized as follows. In the next section (\ref{second}) we review basic facts about PUL expansion of metric and generalize it to the case
of $k-$brane Carrollian limit. Then in section (\ref{third})
we study p-brane in this background and discuss conditions when it is possible to determine its Lagrangian form. 
In section (\ref{fourth}) we generalize this analysis to the case of unstable Dp-brane and solve corresponding equations of motion. 

\section{$k-$Brane Carrollian Limit}\label{second}
In this section we review and generalize small speed of light expansion of general relativity introduced in \cite{Hansen:2021fxi} which was based on  pre-ultra local (PUL) expansion of vielbein. As was shown there the leading order expansion of PUL variables leads to Carroll geometry. Explicitly, in terms of metric $g_{\mu\nu}$ and
its inverse $g^{\mu\nu}$ the small speed of light expansion is the PUL parametrization
\begin{equation}
g_{\mu\nu}=-c^2 T_\mu T_\nu+\Pi_{\mu\nu} \ , \quad 
g^{\mu\nu}=-\frac{1}{c^2}V^\mu V^\nu+\Pi^{\mu\nu}
 \ , 
 \end{equation}
where one form and vector $T_\mu$ and $V^\mu$ were introduced together with 
spatial tensors $\Pi_{\mu\nu}$ and $\Pi^{\mu\nu}$ that obey following
orthogonality and completeness conditions
\begin{equation}\label{TVcon}
T_\mu V^\mu=-1 \ , \quad 
T_\mu \Pi^{\mu\nu}=0 \ , \quad 
\Pi_{\mu\nu}V^\nu=0 \ , \quad  -V^\mu T_\nu+\Pi^{\mu\rho}\Pi_{\rho\nu}=
\delta^\mu_\nu \ .
\end{equation}
This PUL parametrization corresponds to the splitting of tangent space into
temporal and spatial components. 

Few years ago the Carroll limit of strings and p-branes
was studied in \cite{Cardona:2016ytk} where the generalized $k-$brane
and string Carroll limit was introduced. Roughly speaking they correspond to 
taking  Carrollian limit along $(k+1)-$ directions which is generalization 
of standard Carroll limit that is defined as small speed of light expansion  along time-like direction only. 
Following analysis presented in \cite{Hansen:2021fxi} we would like to introduce
PUL parametrization in case of $k-$brane Carroll limit as well. Explicitly, 
we consider following expansion
\begin{equation}\label{expg}
g_{\mu\nu}=c^2 T_\mu^{ \ I}T_\nu^{ \ J}\eta_{IJ}+\Pi_{\mu\nu} \ , 
\quad 
g^{\mu\nu}=\frac{1}{c^2}V^\mu_{ \ I}V^\nu_{ \ J}\eta^{IJ}+\Pi^{\mu\nu} \ , 
\end{equation}
where we have longitudinal forms $T_\mu^{ \ I}, V^\mu_{ \ I}$ with $I=0,1,\dots,k$
and where $\eta_{IJ}=\eta^{IJ}=\mathrm{diag}(-1,1,\dots,1)$

Inserting (\ref{expg}) into  $g_{\mu\nu}g^{\nu\rho}=\delta_\mu^\rho$ we obtain
\begin{equation}
V^\mu_{ \ I}\Pi_{\mu\nu}=0 \ , \quad T_\mu^{ \ I} \Pi^{\mu\nu}=0 \ 
\end{equation}
 and also
\begin{equation}\label{TIT}
T_\mu^{ \ I}T_\nu^{ \ J}\eta_{IJ}V^\nu_{ \ K}V^\rho_{ \ L}\eta^{KL}+
\Pi_{\mu\nu}\Pi^{\nu\rho}=\delta_\mu^\rho \ . 
\end{equation}
To proceed further we generalize first condition in 
(\ref{TVcon}) in the following way
\begin{equation}
T_\mu^{ \ I}V^\mu_{ \ J}=-\delta^I_J \ 
\end{equation} 
so that (\ref{TIT}) takes the form
\begin{equation}
-T_\mu^{ \ I}V^\rho_{ \ I}+\Pi_{\mu\nu}\Pi^{\nu\rho}=\delta_\mu^\rho \ 
\end{equation}
that can be interpreted as generalization of the fourth equation in (\ref{TVcon}).

As the next step we formulate generalized PUL expansion of vielbein components. Recall that the metric can be written as $g_{\mu\nu}=E_\mu^{ \ A}E_\nu^{ \ B}\eta_{AB}, A,B=0,1,\dots,D$ while the inverse metric as $g^{\mu\nu}=
\Theta^\mu_{ \ A}\Theta^\mu_{ \ B}\eta^{AB}$. Then, following \cite{Hansen:2021fxi},  we write
\begin{eqnarray}\label{Edef}
& &E_\mu^{ \ I}=cT_\mu^{ \ I} \ , \quad  I=0,1,\dots,k \ , \quad 
E_\mu^{ \ a}=E_\mu^{ \ a} \ , a=k+1,\dots,D \ , 
\nonumber \\
& &\Theta^\mu_{ \ I}=-\frac{1}{c}V^\mu_{ \ I} \ , \quad 
\Theta^\mu_{ \ a}=\Theta^\mu_{ \ a} \ , \nonumber \\
\end{eqnarray}
where
\begin{equation}
\Pi_{\mu\nu}=E_\mu^{ \ a}\delta_{ab}E_\nu^{ \ b } \ , 
\quad 
\Pi^{\mu\nu}=E^\mu_{ \ a}\delta^{ab}E^\nu_{ \ b} \ . 
\end{equation}
Then by definition we obtain
\begin{eqnarray}
%E_\mu^{ \ A}E^\mu_{ \ B}=\delta_A^B \Rightarrow 
%E_\mu^{  \ I}\Theta^
%\mu_{ \ J}=
T_\mu^{ \ I}\Theta^\mu_{ \ J}=-\delta^I_J \ , \quad 
T_\mu^{ \ I}\Theta^\mu_{ \ a}=0 \ , \quad  E_\mu^{ \ a}V^\mu_{ \ I}=0 \ , 
\quad 
%E_\mu^{ \ I}\Theta^\mu_{ \ a}=-cT_\mu^{ \ I}\Theta^\mu_{ \ a}=0
%\Rightarrow T_\mu^{ \ I}\Theta^\mu_{ \ a}=0 \ , \nonumber \\
%E_\mu^{ \ a}\Theta^\mu_{ \ I}=-E_\mu^{ \ a}V^\mu_{ \ I}= 
%0 \ E_\mu^{ \ a}V^\mu_{ \ I}=0 \ , \nonumber \\
%
E_\mu^{ \ a}\Theta^\mu_{ \ b}=\delta^a_b \ . \nonumber \\
\end{eqnarray}
Let us further presume that these  variables are analytic in $c^2$ so that we can write them as
\begin{eqnarray}\label{expVT}
& &V^\mu_{ \ I}=v^\mu_{ \ I}+c^2 M^\mu_{ \ I} +\mathcal{O}(c^4) \ , \nonumber \\
& &T_\mu^{ \ I}=\tau_\mu^{ \ I}+\mathcal{O}(c^2) \ , \nonumber \\
& &\Theta^\mu_{ \ a}=\theta^\mu_{ \ a}+c^2\pi^\mu_{ \ a}+\mathcal{O}(c^4) \ , \nonumber \\
&&\Pi^{\mu\nu}=h^{\mu\nu}+c^2\Phi^{\mu\nu}+\mathcal{O}(c^4) \  ,\nonumber \\
&&E_\mu^{ \ a}=e_\mu^{ \ a}+\mathcal{O}(c^2) \ , \nonumber \\
&&\Pi_{\mu\nu}=h_{\mu\nu}+\mathcal{O}(c^2) \ . \nonumber \\
\end{eqnarray}
Now inserting expansion (\ref{expVT}) into (\ref{Edef}) we obtain
\begin{eqnarray}
& &\tau_\mu^{ \ I}\theta^\mu_{ \ a}=0 \ , \quad e_\mu^{ \ a}v^\mu_{ \ I}=0 \ , \quad 
\tau_\mu^{ \ I}v^\mu_{ \ J}=-\delta_I^J \ , \nonumber \\
& & e_\mu^{ \ a}\theta^\mu_{ \ b}=\delta^a_b \ , \quad 
-\tau_\mu^{ \ I}v^\nu_{ \ I}+h_{\mu\rho}h^{\rho\nu}=\delta_\mu^\nu \ .
\nonumber \\
\end{eqnarray}
In the same way as in \cite{Hansen:2021fxi} we could study properties of these more general PUL expansions but we  leave it for future work. Instead we focus on the 
analysis of p-branes in these background.
\section{p-Brane in $k-$brane Carrollian Background }\label{third}
In this section we will study p-brane action in $k-$brane Carrollian background.
%Let us introduce  Carroll limit along $k-$ dimensions in case of p-brane action.
 Recall  that p-brane is $p+1$-dimensional object whose action has the form
\begin{equation}
S=-\ttau_p\int d^{p+1}\xi \sqrt{-\det g_{\alpha\beta}} \ ,
\end{equation}
where $g_{\alpha\beta}=g_{\mu\nu}\partial_\alpha x^\mu\partial_\beta x^\nu$ 
is pull-back of target space metric $g_{\mu\nu}$ to $p+1$-dimensional world-volume
of p-brane that is parametrized by coordinates $\xi^\alpha \  , \quad 
\alpha,\beta,\dots=0,1,\dots,p$. Further, $x^\mu(\xi),\mu,\nu,\dots=0,1,\dots,D$ are world-volume fields that define embedding of p-brane into target space-time and
$\partial_\alpha x^\mu\equiv\frac{\partial x^\mu}{\partial \xi^\alpha}$. Finally 
$\ttau_p$ is $p+1$-brane tension.
In order to define $k-$ Carroll limit we consider an PUL expansion of the metric
\begin{equation}
g_{\mu\nu}=c^2\tau_\mu^{ \ I}\tau_\nu^{ \ J}\eta_{IJ}+h_{\mu\nu} \ , 
\end{equation}
where $I,J=0,1,\dots,k$. 
 Then the induced
metric has the form
\begin{equation}
g_{\alpha\beta}=c^2 \tau_\alpha^{ \ I}\tau_\beta^{ \ J}\eta_{IJ}+h_{\alpha\beta} \ . 
\end{equation}
As the next step we introduce canonical form of p-brane action
\begin{equation}\label{Scanon}
S=\int d^{p+1}\xi (p_\mu \partial_0 x^\mu-\tN\mH_\tau-\tN^i\mH_i) \ , 
\end{equation}
where $\mH_\tau$ and $\mH_i$ are following first class constraints
\begin{eqnarray}
& &\mH_\tau=p_\mu g^{\mu\nu}p_\nu+\ttau_p^2\det g_{ij} \ , \quad   g_{ij}=\partial_i x^\mu g_{\mu\nu}\partial_j x^\nu \ , \nonumber \\
& & \mH_i=p_\mu\partial_i x^\mu  \ , \quad i,j=1,\dots,p \ . \nonumber \\
\end{eqnarray}
Now we are ready to proceed to the formulation of p-brane action in $k-$brane Carrollian background. Note that  the PUL expansion of the inverse metric has the form
\begin{equation}\label{ginv}
g^{\mu\nu}=\frac{1}{c^2}V^\mu_{ \ I}V^\nu_{ \ J}\eta^{IJ}+\Pi^{\mu\nu} \ . 
\end{equation}
Due to the prefactor $c^{-2}$ (\ref{ginv}) we see that the action 
(\ref{Scanon}) is finite when we rescale $\tN$ as
\begin{equation}
\tN=c^2N \ . 
\end{equation}
However then it is also clear from the  Hamiltonian constraint 
that we should rescale $\ttau_p$ as
\begin{equation}
\ttau_p=\frac{1}{c}\tau_p 
\end{equation}
while we do not rescale $\tN^i$ so that $\tN^i=N^i$.
%Then we also rescale $\tN$ and $\ttau_p$ as
%\begin{equation}
%\tN=c^2N \ , \quad \tN^i=\tN^i \ , \quad \ttau_p=\frac{1}{c}\tau_p
%\end{equation}
As a result we obtain Dp-brane action in the $k-$Carrollian background
\begin{eqnarray}
S
%=\int d^{p+1}\xi 
%(p_\mu\partial_0 x^\mu-N(p_\mu V^\mu_{ \ I}V^\nu_{ \ J}\eta^{IJ}p_\nu
%+\tau_p^2\det h_{ij})-N^ip_\mu \partial_i x^\mu)=\nonumber \\
=\int d^{p+1}\xi
(p_\mu\partial_0 x^\mu-N(p_\mu v^\mu_{ \ I}\eta^{IJ}
v^\nu_{ \ J}p_\nu+\tau_p^2\det h_{ij})-N^ip_\mu\partial_i x^\mu) \ ,
\nonumber \\
\end{eqnarray}
where $h_{ij}=h_{\mu\nu}\partial_i x^\mu \partial_j x^\nu$. Note that this canonical form of p-brane action can be formulated in any $k-$brane Carrollian background.

Let us now determine corresponding Lagrangian density. Note that the Hamiltonian has the form
\begin{equation}
H=\int d^p\xi \mH \ , \quad \mH=N (p_\mu v^\mu_{ \ I}\eta^{IJ}
v^\nu_{ \ J}p_\nu+\tau_p^2\det  h_{ij})+N^ip_\mu \partial_i x^\mu \ .
\end{equation}
Using this Hamiltonian we obtain 
\begin{equation}
\partial_0 x^\mu=\pb{x^\mu,H}=2Nv^\mu_{ \ I}\eta^{IJ}
v^\nu_{ \ J}p_\nu+N^i\partial_i x^\mu \ 
\end{equation}
and hence Lagrangian density is equal to
\begin{eqnarray}\label{LNNi}
& &\mL=p_\mu\partial_0 x^\mu-\mH=
N p_\mu v^\mu_{ \ I}\eta^{IJ}v^\nu_{ \ J}p_\nu-N\tau_p^2\det h_{ij}=
\nonumber \\
& &=\frac{1}{4N}(\partial_0 x^\mu-N^i\partial_i x^\mu)\tau_\mu^{ \ I}\eta_{IJ}
\tau_\nu^{ \ J}(\partial_0 x^\nu-N^i\partial_i x^\nu)-N\tau_p^2\det  h_{ij} \ , \nonumber \\
\end{eqnarray}
using
\begin{eqnarray}
%\tau_\mu^{ \ K}(\partial_0 x^\mu-N^i\partial_i x^\mu)=
%-N \eta^{KJ}v^\nu_{ \ J}p_\nu
%\Rightarrow 
%\nonumber \\
v^\nu_{ \ J} p_\nu=-\frac{1}{2N}\eta_{JK}\tau_\mu^{ \ K}
(\partial_0 x^\mu-N^i\partial_i x^\mu) \ .  \nonumber \\
\end{eqnarray}
Finally we should solve the equations of motion for $N^i,N$. For $N^i$ it has the form
\begin{equation}
\tau_{i0}-\tau_{ij}N^j=0 \ , 
\end{equation}%\partial_i x^\mu\tau_\mu^{ \ I}\eta_{IJ}\tau_\nu^{ \ J}
%(\partial_0 x^\nu-N^i\partial_i x^\nu)=0 \Rightarrow 
where we defined $\tau_{\alpha\beta}$ as
\begin{equation}
\tau_{\alpha\beta}=\tau_\alpha^{ \ I}\eta_{IJ}\tau_\beta^{ \ J} \ . 
\end{equation}
Next step depends on the  question  whether $\tau_{ij}$ is non-singular matrix or not. 
In this section we will presume that  $\tau_{ij}$ is non-singular matrix and denote its inverse as
$\ttau^{ij}$ so that 
\begin{equation}
\tau_{ij}\ttau^{jk}=\delta_i^k \ . 
\end{equation}
Then $N^i$ is equal to
\begin{equation}\label{Nisol}
N^i=\ttau^{ij}\tau_{j0} \ . 
\end{equation}
Further, the equation of motion for $N$ has the form
\begin{equation}
-\frac{1}{4N^2}
(\tau_{00}-2N^i\tau_{i0}+\tau_{ij}N^iN^j)-\tau_p^2\det h_{ij}=0 \ .
\end{equation}
Inserting (\ref{Nisol}) into the equation above we find equation for $N$ that
has solution
\begin{eqnarray}\label{Nsol}
%-\frac{1}{N^2}
%(\tau_{00}-2N^i\tau_{i0}+\tau_{ij}N^iN^j)-\tau_p^2\det h_{ij}=0 \Rightarrow 
%\nonumber \\
%-\frac{1}{N^2}(\tau_{00}-\tau_{0i}\ttau^{ij}\tau_{j0})-\tau_p^2\det h_{ij} 
%\Rightarrow \nonumber \\
%-\frac{1}{N^2}\frac{\det \tau_{\alpha\beta}}{\det \tau_{ij}}=\tau_p^2\det h_{ij} \Rightarrow 
N=\frac{1}{2\tau_p}\sqrt{-\frac{\det \tau_{\alpha\beta}}{\det  h_{ij}\det \tau_{ij}}} \ .  \nonumber \\
\end{eqnarray} 
Finally inserting (\ref{Nisol}) and (\ref{Nsol}) into Lagrangian density (\ref{LNNi}) we finally get
\begin{eqnarray}\label{mLCarrnon}
\mL
%=-\tau_p \sqrt{-\det \tau_{\alpha\beta}}\sqrt{\frac{\det h_{ij}}{\det \tau_{ij}}}
%	-\tau_p \sqrt{-\det \tau_{\alpha\beta}}\sqrt{\frac{\det \tau_{\alpha\beta}}{\det h_{ij}}}=\nonumber \\
=	-\tau_p \sqrt{-\det \tau_{\alpha\beta}}\sqrt{\frac{\det h_{ij}}{\det \tau_{ij}}} \ . 
	\nonumber \\	
	\end{eqnarray}
	This is final form of the Lagrangian density for Dp-brane in 
	$k-$Carroll background that was derived on presumption that matrices
	$\tau_{ij}$ and $h_{ij}$ are non-singular. Let us discuss these conditions in more details. 
	
As the first case we discuss the special situation when $k=0$.
\subsection{$k=0$}
For $k=0$  the equation of motion for $N^i$ has the form
	\begin{equation}
-\tau_i\tau_0+\tau_i\tau_j N^j=0
\end{equation}
that for non-zero $\tau_i$ implies 
\begin{equation}
\tau_i N^i=\tau_0 \ . 
\end{equation}	
Then inserting back to the Lagrangian density we get
\begin{eqnarray}
\mL=\frac{1}{N}(-\tau_0^2+2N^i\tau_i\tau_0-N^i\tau_iN^j\tau_j)-N\tau_p^2\det h_{ij}=-N\tau_p^2\det h_{ij}\nonumber \\
\end{eqnarray}
and we see that there is no dynamics at all which is in agreement with the
particle Carroll limit studied in \cite{Cardona:2016ytk}.
	
\subsection{General $k$}
Now we discuss in more details conditions under the matrices  
$\tau_{ij}$ and $h_{ij}$ are non-singular.  Recall that $\tau_{ij}=
\tau_i^{ \ I}\eta_{IJ}\tau_j^{ \ J}$. Now $\tau_i^{ \ I}$ is $p\times (k+1)$ matrix that has the rank $\mathrm{min}(p,k+1)$. Further, $\eta_{IJ}$ is the matrix of the rank $k+1$. Then $\tau_{ij}$, which is $p\times p$ matrix has either rank
 $k+1$(for $k+1<p$) or $p$(for $p<k+1$). Let us discuss these two possibilities in more details.
\begin{itemize}
	\item
{\bf $\mathbf{k+1<p}$}

Then the matrix $\tau_{ij}$ is singular and we cannot solve equations of motion for $N^i$. Further, $h_{ij}=e_i^{ \ a}e_j^{ \ b}\delta_{ab}$ where $e_i^{ \ a}$ is $p\times (D-k)$ matrix
that is non-singular for $p\leq D-k$. If we combine these two conditions together we get
$2k<D-1$ and hence $k<\frac{D-1}{2}$. On the other hand the matrix $h_{ij}$ is singular for $p>D-k$. In this case the Lagrangian density takes the form
\begin{equation}
\mL
=\frac{1}{N}(\partial_0 x^\mu-N^i\partial_i x^\mu)\tau_\mu^{ \ I}\eta_{IJ}
\tau_\nu^{ \ J}(\partial_0 x^\nu-N^i\partial_i x^\nu) \nonumber \\
\end{equation}
that has effectively the form of massless Carroll Dp-brane. For example, in case of the string Carroll limit when $k=1$ and when $D=9$ we get that this corresponds to massless D8-brane. 
%{$\mathbf{K=0}$}
%
%In this case the matrix $h_{ij}$ is always non-singular while matrix $\tau_{ij}$ is 
%singular for $p>1$.In fact, in this case the equation of motion for $N^i$ has the form
%\begin{equation}
%-\tau_i \tau_0+\tau_i\tau_j N^j=0
%\end{equation}
%that implies 
%\begin{equation}
%\tau_0=\tau_j N^j
%\end{equation}
%Note also that Lagrangian densitz has the form 
%\begin{eqnarray}
%\mL=\frac{1}{N}(-\tau_0^2+2N^i\tau_i\tau_0-N^i\tau_iN^j\tau_j)-N\tau_p^2\det h_{ij}=-N\tau_p^2\det h_{ij}\nonumber \\
%\end{eqnarray}
%and we see that there is no dynamics at all. 
%Let us consider second case when $K=1$
%{$\mathbf{K=1}$}
\item $\mathbf{k+1\geq p}$

In this case the matrix $\tau_{ij}$ is non-singular. Further, $h_{ij}$ is non-singular
for $p\leq (D-K)$.  This is exactly situation that leads to the Lagrangian density
(\ref{mLCarrnon}). On the other 
hand when $D-K<p$ we find that $h_{ij}$ is singular and Lagrangian density has the form
\begin{equation}\label{mLhsin}
\mL=-\frac{1}{N}
(\partial_0 x^\mu-N^i\partial_i x^\mu)\tau_\mu^{ \ I}\eta_{IJ}
\tau_\nu^{ \ J}(\partial_0 x^\nu-N^i\partial_i x^\nu) \ . 
\end{equation}
Now we can solve equations of motion for $N^i$ that gives
\begin{equation}
\tau_{0i}=N^j
\tau_{ji}
\end{equation}
that can be solved for $N^j$ introducing inverse matrix $\ttau^{ij}$. Inserting back to the Lagrangian density (\ref{mLhsin}) we obtain 
\begin{equation}
\mL=-\frac{1}{N}(\tau_{00}-\tau_{0i}\ttau^{ij}\tau_{j0})=
-\frac{1}{N}\frac{\det \tau_{\alpha\beta}}{\det \tau_{ij}} \ . 
\end{equation}
From this Lagrangian density we immediately see that the dynamics is trivial
since the equation of motion for $N$ implies that $\det \tau_{\alpha\beta}=0$ that
is satisfied when $\partial_0 x^\mu=0$ due to the fact that $\tau_{ij}$ is non-singular.

\end{itemize}

\section{Carroll Limit of Non-BPS Dp-Brane}\label{fourth}
In this section we consider non-BPS Dp-brane in 
$k-$brane Carrollian background.  For simplicity we will consider the case of zero NSNS two form and RR background. In this case the tachyon effective action has the form
\begin{equation}\label{SDBInon}
S=-\ttau_{non,p} \int d^{p+1}e^{-\phi}V(T)\sqrt{-\det (g_{\alpha\beta}+\tl_s^2 \tF_{\alpha\beta}+\partial_\alpha T\partial_\beta T)} \ , 
\end{equation}
where 
\begin{equation}
g_{\alpha\beta}=g_{\mu\nu}\partial_\alpha x^\mu\partial_\beta x^\nu
\ , \quad \tF_{\alpha\beta}=\partial_\alpha \tA_\beta-\partial_\beta \tA_\alpha \ . 
\end{equation}
Further, $T$ is the tachyon field defined on the world-volume of unstable Dp-brane,
$V(T)$ is its potential when we presume that $V$ is even function with three extremes, where $T_0 = 0$
is unstable maximum with $V (T_0)= V_{max}$ while $\pm \infty$ are global minima of the potential with $V (\pm \infty) = 0$. Finally $\ttau_{non,p}$ is tension for non-BPS Dp-brane and $\phi$ is dilaton which generally depends on $x^\mu$.

As in previous section we begin with the  canonical form of the action (\ref{SDBInon}) that has the form
\begin{eqnarray}
& &S=\int d^{p+1}\xi (p_\mu\partial_0 x^\mu+\tpi^i\partial_0 \tA_i-\nonumber \\
& &-\tN(p_\mu g^{\mu\nu}p_\nu+p_T^2+\tpi^i(g_{ij}+\partial_i T\partial_j T)\tpi^j+\ttau_{non,p}^2V^2 e^{-2\phi}\det (g_{ij}+\tl_s^2 \tF_{ij}+\partial_i T\partial_j T))-
\nonumber \\
& &-\tN^i(p_\mu\partial_0 x^\mu+\tF_{ij}\tpi^j+p_T\partial_iT)-\tpi^i\partial_i \tA_0) \ , \nonumber \\
\end{eqnarray}
where $p_T$ is momentum conjugate to $T$ and where $\tpi^i$ is momentum conjugate to $\tA_i$. 
Let us again consider metric components in the form
\begin{equation}
g_{\mu\nu}=c^2\tau_\mu^{ \ I}\tau_\nu^{ \ J}\eta_{IJ}+h_{\mu\nu} \ , 
\quad 
g^{\mu\nu}=\frac{1}{c^2}V^\mu_{ \ I}V^\nu_{ \ J}\eta^{IJ}+\Pi^{\mu\nu} \ , 
\end{equation}
so that the action has the form
\begin{eqnarray}
& &S=\int d^{p+1}\xi (p_\mu\partial_0 x^\mu+\tpi^i\partial_0 \tA_i-\nonumber \\
& &-\tN(p_\mu (\frac{1}{c^2}V^\mu_{ \ I}V^\nu_{ \ J}\eta^{IJ}+\Pi^{\mu\nu})p_\nu+\tpi^i
(c^2\tau_i^{ \ I}\tau_j^{ \ J}\eta_{IJ}+h_{ij}+\partial_i T\partial_j T)
\tpi^j+p_T^2\nonumber \\
& &+\ttau_{non,p}^2 e^{-2\phi}V^2\det (c^2\tau_i^{ \ I}\tau_j^{ \ J}\eta_{IJ}+h_{ij}+\tl_s^2 \tF_{ij}+\partial_i T\partial_j T))-
\nonumber \\
& &-\tN^i(p_\mu\partial_0 x^\mu+\tF_{ij}\tpi^j+p_T\partial_i T)-\tpi^i\partial_i \tA_0) \ . \nonumber \\
\end{eqnarray}
First of all we see that we have to scale $\tN$ as 
\begin{equation}
\tN=c^2 N
\end{equation}
in order to keep action finite. Then however to keep non-trivial gauge field we should scale the momenta 
$\tpi^i$ as 
\begin{equation}
\tpi^i=\frac{1}{c}\pi^i \ , \quad \tA_\alpha=c A_\alpha \ . 
\end{equation}
Finally if we rescale $\ttau_p$ as $\ttau_{non,p}=\frac{1}{c}\tau_{non,p}$ we should be careful with rescaling $\tl_s$. Recall that $\ttau_{non,p}\sim \tl_s^{-(p+1)}$ so that we should scale $\tl_s$ as 
\begin{equation}
%\tl_s^{p+1}=cl_s^{p+1} \Rightarrow 
\tl_s=c^{\frac{1}{p+1}}l_s \ . 
\end{equation}
Then in the limit $c\rightarrow 0$ we find that the action takes the form
\begin{eqnarray}\label{Scannon}
& &S=\int d^{p+1}\xi (p_\mu\partial_0 x^\mu+\pi^i\partial_0 A_i+p_T\partial_0 T-\nonumber \\
& &-N(p_\mu v^\mu_{ \ I}\eta^{IJ}v^\nu_{ \ J}p_\nu+\pi^i(
h_{ij}+\partial_iT\partial_jT)\pi^j
+\tau_{non,p}^2 e^{-2\phi}V^2\det (h_{ij}+\partial_i T\partial_jT))-
\nonumber \\
& &-N^i(p_\mu\partial_i x^\mu+F_{ij}\pi^j+p_T\partial_i T)-\pi^i\partial_i A_0) \ . \nonumber \\
\end{eqnarray}
This is final form of non-BPS Dp-brane in $k-$brane Carrollian background. 
In the next section we  focus on the canonical equations of motion and discuss their solutions.
\subsection{Equations of Motion and Their Solutions}
Now we determine equations of motion that follow from the Hamiltonian density 
of the action (\ref{Scannon}) that has the form
\begin{eqnarray}
& &H=\int d^p\xi \mH \ , \quad 
\mH=N\mH_\tau+N^i\mH_i-\pi^i\partial_i A_0 \ , \quad \mH_i=p_\mu\partial_i x^\mu+F_{ij}\pi^j+p_T\partial_i T \ , \nonumber \\
& &\mH_\tau=p_\mu v^\mu_{ \ I}\eta^{IJ}v^\nu_{ \ J}p_\nu+\pi^i(
h_{ij}+\partial_iT\partial_jT)\pi^j
+\tau_{non,p}^2 e^{-2\phi}V^2\det (h_{ij}+\partial_i T\partial_jT) \ .
\nonumber \\
\end{eqnarray}
With the help of this Hamiltonian we obtain following canonical equations of motion
\begin{eqnarray}\label{nonBPSeom}
& &\partial_0 T=\pb{T,H}=N^i\partial_i T \ , \nonumber \\
& &\partial_0 p_T=\pb{p_T,H}=-2N\tau_{non,p}^2 e^{-2\phi}V\frac{dV}{dT}\det (h_{ij}+N\partial_iT \partial_jT)+\nonumber \\
& &+2\tau_{non,p}^2 \partial_i[N e^{-2\phi}V^2(\bA^{-1})^{ij}\partial_j T
\det \bA]+\partial_i[N^i p_T] 
+2\partial_i[N \pi^j\partial_j T\pi^i]
\ , \nonumber \\
& &\partial_0 x^\mu=\pb{x^\mu,H}=
2Nv^\mu_{ \ I}v^\nu_{ \ J}\eta^{IJ}p_\nu+N^i\partial_i x^\mu \ , 
\nonumber \\
& &\partial_0 A_i=\pb{A_i,H}=
2Nh_{ij}\pi^j-F_{ij}N^j+\partial_i A_0 \ , \nonumber \\
& &\partial_0 p_\mu=\pb{p_\mu,H}=
-Np_\rho \partial_\mu (v^\mu_{ \ I}v^\nu_{ \ J}\eta^{IJ})
p_\nu -\partial_i[N^i p_\mu]\nonumber \\
& &+2\partial_i[N \pi^ih_{\mu\nu}\partial_j x^\nu \pi^j]
-N\pi^i\partial_i x^\rho\partial_\mu h_{\rho\sigma}
\partial_j x^\sigma -\tau_{non,p}^2\partial_\mu[e^{-2\phi}]V^2
\det(h_{ij}+\partial_i T\partial_j T)+\nonumber \\
& &+2\tau_{non,p}^2 \partial_i[N e^{-2\phi}V^2 h_{\mu\nu}\partial_j x^\nu \bAi^{ji}
\det \bA_{ij}]-
\tau_{non,p}^2 e^{-2\phi}NV^2\partial_i x^\rho \partial_\mu h_{\rho\sigma}
\partial_j x^\sigma \bAi^{ji}\det \bA \ , \nonumber \\
& &\partial_0 \pi^i=\pb{\pi^i,H}=\partial_k[N^k\pi^i]-\partial_l[N^i\pi^l] \ ,
\nonumber \\
\end{eqnarray}
where we introduced matrix $\bA_{ij}=h_{ij}+\partial_i T\partial_j T$ and $\bAi^{jk}$ is its inverse $\bA_{ij}\bAi^{jk}=\delta_i^k$.

Let us  solve the equations of motion 
(\ref{nonBPSeom})
 with the ansatz corresponding to the tachyon kink solution. Note that the tachyon kink solution was previously studied in the context of relativistic non-BPS Dp-branes
in \cite{Sen:2003tm,Kluson:2016tgp,Kluson:2005fj,Kluson:2005hd}, for review and extensive list of references, see \cite{Sen:2004nf}. We apply similar procedure in case
of non-BPS Dp-brane in $k-$ brane Carrollian background where however we solve canonical
equations of motion (\ref{nonBPSeom}) instead of Lagrangian ones that were analysed in 
\cite{Sen:2003tm,Kluson:2016tgp,Kluson:2005fj,Kluson:2005hd}.

  Explicitly, we presume that the tachyon has the profile
\begin{equation}\label{fans}
T=f(a(y-t(\xi^{\hi}))) \ , 
\end{equation}
where we select one world-volume coordinate $\xi^p=y$ and where 
$\hi=1,\dots,p-1$. Note that we presume that $T$ does not depend on $\xi^0$. We further presume that all world-volume fields 
depend on $\xi^{\halpha}$ only so that  
\begin{equation}\label{ansxA}
x^\mu(\xi^\alpha)=\tx^\mu(\xi^{\halpha}) \ , \quad A_{\hi}(\xi^\alpha)=\ta_{\hi}(\xi^{\halpha}) \ , \quad A_y=0 \ , \quad 
A_0(\xi^{\alpha})=\ta_0(\xi^{\halpha}) \  ,
\end{equation}
where $\halpha=0,1,\dots,p-1$.
The function $f(u)$ that appears in (\ref{fans}) satisfies conditions
\begin{equation}
f(u)=-f(-u) \ , \quad  f'(u)>0 \ \forall \  u \ , \quad f(\pm \infty)=\pm \infty
\end{equation}
but is otherwise an arbitrary function of its argument \cite{Sen:2003tm}.

From the reasons that will be clear below we suggest following ansatz
for conjugate momenta
\begin{equation}\label{pmuans}
p_\mu(\xi^\alpha)=\frac{\tau_{non,p}}{\tau_{p-1}}af'V\tp_\mu(\xi^{\halpha}) \ , \quad \pi^{\hi}(\xi^\alpha)=\frac{\tau_{non,p}}{\tau_{p-1}}af'V\tpi^{\hi}(\xi^{\halpha}) \ ,
\end{equation}
where the factor $\frac{\tau_{non,p}}{\tau_{p-1}}$ was included through dimensional reasons. 
We also presume that $N$ has the form
\begin{equation}\label{Npres}
N(\xi^\alpha)=\frac{\tau_{p-1}}{\tau_{non,p}}\tn(\xi^{\halpha})\frac{1}{af'V} \ ,  
\end{equation}
where again the factor $\frac{\tau_{p-1}}{\tau_{non,p}}$ was include through dimensional reasons. 

Inserting (\ref{fans}) into equation of motion for $T$ we get
\begin{equation}
0=
N^y af'-N^{\hi}\partial_{\hi}t f'a
 \nonumber \\
\end{equation}
that can be solved for $N^y$ as
\begin{equation}\label{Nysol}
N^y=\tn^{\hi}\partial_{\hi}t \  , \quad N^{\hi}(\xi^{\alpha})\equiv \tn^{\hi}(\xi^{\halpha}) \ .
\end{equation}
 Further, the equation of motion for $N^y$ has the form
\begin{equation}
\partial_y x^\mu p_\mu+\partial_y T p_T=0
\end{equation}
that, since $\partial_y T\neq 0$ and since $\partial_y x^\mu=0$ for the ansatz (\ref{ansxA})
 implies that $p_T=0$. 

Let us further consider equation of motion for $\pi^{\hi}$ that for the ansatz
(\ref{ansxA}) and (\ref{pmuans}) has the form
\begin{eqnarray}\label{phieqm}
& &af'V\partial_0\tpi^{\hi}
%=\partial_k[N^k\pi^{\hi}]-\partial_l[N^{\hi}\pi^{l}]
%\partial_y[N^y af'V\tpi^{\hi}]+\partial_{\hk}[N^{\hk}af'V\tpi^{hi}]-
%\partial_{\hj}[N^{\hi}af'V\tpi^{\hj}]-
%\partial_y[N^{\hi}\pi^y]=\nonumber \\
=\partial_y[af'V]N^y\tpi^{\hi}+\partial_{\hk}[af'V]N^{\hk}
\tpi^{\hi}+af'V\partial_{\hk}[N^{\hk}\tpi^{\hi}]-\nonumber \\
& &-\partial_{\hj}[af'V]\tpi^{\hj}N^{\hi}-af'V\partial_{\hj}[N^{\hi}
\tpi^{\hj}]-\partial_y \pi^y N^{\hi}=\nonumber \\
& &=
%[\partial_y[af'V]N^{\hj}\partial_{\hj}t-\partial_y[af'V]\partial_{\hk}tN^{\hk}]\tpi^{\hi}
af'V\left(\partial_{\hk}[\tn^{\hk}\tpi^{\hi}]
-\partial_{\hj}[\tn^{\hi}\tpi^{\hj}]\right)+
\partial_y[af'V]\partial_{\hj}t \tpi^{\hj}\tn^i-\partial_y\pi^y\tn^i \ , 
\nonumber \\
\end{eqnarray}
where we used (\ref{Nysol}) together with the fact that
\begin{equation}\label{phihiafT}
\partial_{\hi}(af'V)=-\partial_y (af'V)\partial_{\hi}t \ 
\end{equation}
as follows from the form of the tachyon field given in (\ref{fans}). Now in order to cancel
term proportional to $\partial_y[af'V]$ we suggest that $\pi^y$ is equal to
\begin{equation}\label{pyans}
\pi^y=af'V\tpi^{\hi}\partial_{\hi}t \ . 
\end{equation}
With this ansatz the equation (\ref{phieqm}) reduces into
\begin{equation}
af'V\left[\partial_0\tpi^{\hi}-
\left(\partial_{\hk}[\tn^{\hk}\tpi^{\hi}]
-\partial_{\hj}[\tn^{\hi}\tpi^{\hj}]\right)\right]=0
\ . 
\end{equation}
Now $af'V$ goes to zero in the limit $a\rightarrow \infty$ for $y\neq t(\xi^{\hi})$. This can be easily seen when we note that $V(T)\sim e^{-T}$ for large $T$
\cite{Sen:2003tm}. On the other hand for $y=t(\xi^{\hi})$ we get that $af'V$ is finite at $y=t(\xi^{\hi})$ and hence in order to obey equation of motion for $\pi^i$ we have to impose condition that
\begin{equation}
\partial_0\tpi^{\hi}-
\partial_{\hk}[\tn^{\hk}\tpi^{\hi}]
+\partial_{\hj}[\tn^{\hi}\tpi^{\hj}]=0
\end{equation}
that are equations of motion for $\tpi^{\hi}$ on the world-volume of D(p-1)-brane in 
$k-$brane Carrollian background. In more details, following limiting procedure performed in case of non-BPS Dp-brane we obtain canonical action for D(p-1)-brane  in $k-$brane Carrollian background in the form 
\begin{eqnarray}\label{SactDp}
& &S=\int d^{p-1}\xi (\tp_\mu\partial_0\tx^\mu+\tpi^{\hi}\partial_0 \ta_{\hi}-\tn \tmH-
\tn^{\hi}\tmH_i-\tpi^{\hi}\partial_{\hi} \ta_0) \ , \nonumber \\
& &\tmH=
\tp_\mu v^\mu_{ \ I}\eta^{IJ}v^\nu_{ \ J}\tp_\nu+\tpi^{\hi}
h_{\hi\hj}\tpi^{\hj}
+\tau_{p-1}^2 e^{-2\phi}\det h_{\hi\hj} \ , \quad 
\tmH_{\hi}=\tp_\mu\partial_{\hi}\tx^\mu+\tF_{\hi\hj}\tpi^{\hj}\ . \nonumber \\
\end{eqnarray}
The equations of motion that follow from this action have the form 
\begin{equation}\label{eqxp}
\partial_0 \tx^\mu=
2\tn v^\mu_{ \ I}v^\nu_{ \ J}\eta^{IJ}\tp_\nu+\tn^{\hi}\partial_{\hi} \tx^\mu \ , 
\end{equation}
\begin{equation}\label{eqap}
\partial_0 \ta_{\hi}=
2\tn h_{\hi\hj}\tpi^{\hj}-\tF_{\hi\hj}\tn^j+\partial_{\hi} \ta_0 \ , 
\end{equation}
and
\begin{eqnarray}\label{eqpmup}
& &\partial_0 \tp_\mu=
-\tn\tp_\rho \partial_\mu (v^\mu_{ \ I}v^\nu_{ \ J}\eta^{IJ})
p_\nu -\partial_{\hi}[\tn^{\hi} p_\mu]\nonumber \\
& &+2\partial_{\hi}[\tn \tpi^{\hi}h_{\mu\nu}\partial_{\hj} \tx^\nu \tpi^{\hj}]
-\tn\tpi^{\hi}\partial_{\hi} \tx^\rho\partial_\mu h_{\rho\sigma}
\partial_{\hj} \tx^\sigma -\tau_{p-1}^2\partial_\mu[e^{-2\phi}]
\det h_{\hi\hj}+\nonumber \\
& &+2\tau_{p-1}^2 \partial_{\hi}[\tn e^{-2\phi} h_{\mu\nu}\partial_{\hj} \tx^\nu h^{\hj\hi}
\det h_{\hi\hj}]-\tau_{p-1}^2 e^{-2\phi}\tn \partial_{\hi}\tx^\rho
\partial_\mu h_{\rho\sigma}\partial_{\hj}\tx^\sigma
h^{\hj \hi}\det h_{\hi\hj}
 \ , \nonumber \\
\end{eqnarray}
and finally
\begin{equation}\label{eqapi}
\partial_0 \tpi^{\hi}=\partial_{\hk}[\tn^{\hk}\tpi^{\hi}]-\partial_{\hk}[\tn^{\hi}\tpi^{\hk}] \ .
\end{equation}
We see that the equation (\ref{phieqm}) is obeyed when the conjugate momenta
$\tpi^{\hi}$ obey the equations of motion (\ref{eqapi}). In other words, the dynamics of conjugate momenta $\tpi^{\hi}$ at the core of  kink  localized at $y=t(\xi^{\hi})$ is governed equations of motion that follow from the variation of D(p-1)-brane action in $k-$brane Carrollian background. 

As the next step
we should study equation of motion for $\pi^y$. Firstly inserting  
(\ref{pmuans}) and (\ref{pyans}) together with (\ref{Nysol}) into right side of the equation of motion for $\pi^y$ we obtain
\begin{eqnarray}\label{timepi0r}
%\partial_0\pi^y=\partial_k[N^k\pi^y]-\partial_l[N^y\pi^l]=
%\partial_{\hk}[\tn^{\hk} af'V\tpi^{\hi}\partial_{\hi}t]-
%\partial_{\hk}[\tn^{\hat{l}}\partial_{\hat{l}}t af'V\tpi^{\hk}]=\nonumber \\
%=af'V(\partial_{\hk}[\tn^{\hk}\pi^{\hi}]-\partial_{\hk}[\tpi^{\hk}\tn^{\hi}])\partial_{\hi}t+
%af'V[\tn^{\hk}\tpi^{\hi}\partial_{\hk}\partial_{\hi}t-
%\partial_{\hk}\partial_{\hat{l}}t\tn^{\hat{l}}\tpi^{\hk}]+
%\nonumber \\
%-\partial_{y}[af'V]\partial_{\hk}t\tn^{\hk}\tpi^{\hi}\partial_{\hi}t+
%\partial_{y}[af'V]\partial_{\hk}t\tn^{\hat{l}}\partial_{\hat{l}}t\tpi^{\hk}=\nonumber \\
af'V(\partial_{\hk}[\tn^{\hk}\pi^{\hi}]-\partial_{\hk}[\tpi^{\hk}\tn^{\hi}])\partial_{\hi}t \ , 
\end{eqnarray}
where we also used (\ref{phihiafT}). On the other hand time derivative of 
(\ref{pyans}) gives
\begin{equation}\label{timepi0}
\partial_0\pi^y=af'V\partial_0\tpi^{\hi}\partial_{\hi}t \ . 
\end{equation}
If we combine (\ref{timepi0}) with (\ref{timepi0r})
we get that the equation of motion for $\pi^y$ is obeyed on condition when 
$\tpi^{\hi}$ solves the equations of motion (\ref{eqapi}) which proves consistency
of our ansatz.

As the next step we proceed to the equations of motion for $T$ and $p_T$. Note that
for the ansatz (\ref{fans}) and (\ref{ansxA}) we have
\begin{equation}\label{detans}
\det (h_{ij}+\partial_i T\partial_j T)=
\det h_{\hi\hj}a^2f'^2 \ .
\end{equation}
Then after some calculations we find that right side of the equation of motion for $p_T$ is identically zero 
\begin{eqnarray}
%0=-2N\tau_p^2 V\frac{dV}{dT}a^2f'^2\det h_{\hi\hj}+2\tau_p^2 \partial_i[N e^{-2\phi}
%V^2\bAi^{ij}\partial_j T\det \bA]+2\partial_i[\pi^i \partial_j T\pi^j]
%\nonumber \\
%=-2\tn\tau_p^2 e^{-2\phi}\frac{dV}{dT}af'\det h_{\hi\hj}+
%+2\tau_p^2\partial_i[\tn e^{-2\phi}V af' \bAi^{ij}\partial_j T
%\det h_{\hi\hj}]=\nonumber \\
%4\tau_p^2 V\frac{dV}{dT}e^{-2\phi}N\partial_i T
%\bAi^{ij}\partial_j T\det \bA+\nonumber \\
%+2\tau_p^2V^2 \partial_{\hi}[e^{_2\phi}N]\bAi^{\hi j}\partial_j T
%\det \bA+
%2\tau_p^2e^{-2\phi}NV^2\partial_i[\bAi^{ij}\partial_j T\det \bA]=\nonumber \\
& &-2\tn e^{-2\phi}\frac{dV}{dT}af'\det h_{\hi\hj}+\nonumber \\
&&\partial_y[\tn e^{-2\phi} V^2a^2f'^2(\frac{1}{a^2f'^2}+\partial_{\hat{m}}th^{\hat{m}
	\hat{n}}\partial_{\hat{n}}t)\det h_{\hi\hj}]-\partial_y
[\tn e^{-2\phi}V^2 a^2f'^2\partial_{\hat{m}}
t h^{\hat{m}\hi}\partial_{\hi}t\det h_{\hi\hj}]+\nonumber \\
& &+\partial_{\hi}[\tn e^{-2\phi}V^2 a^2f'^2\partial_{\hat{m}}t
h^{\hat{m}\hi}\det h_{\hi\hj}]-\partial_{\hi}[\tn e^{-2\phi}V^2a^2f'^2
h^{\hi \hj}\partial_{\hj}t\det h_{\hi\hj}]=0
\nonumber \\
\end{eqnarray}
for the ansatz (\ref{fans}) and (\ref{ansxA}). In more details, we used the fact that the matrix  has the form
\begin{equation}
\bAi=\left(\begin{array}{cc}
h^{\hi\hj} &  h^{\hi \hat{m}}\partial_{\hat{m}}t \\
\partial_{\hat{m}} th^{\hat{m}\hj} & \frac{1}{a^2 f'^2}+\partial_{\hat{m}}t
h^{\hat{m}\hat{n}}\partial_{\hat{n}}t \\
\end{array}
\right)
\end{equation}
which, together with (\ref{phihiafT}) also implies
\begin{eqnarray}
& &\partial_i[Ne^{-2\phi}V^2 \bAi^{ij}\partial_jT \det \bA_{ij}]
%\partial_y[\tn e^{-2\phi}Vaf' \bAi^{yy}\partial_y T\det h_{\hi\hj}]+
%\nonumber \\
%\partial_y[\tn e^{-2\phi}V af'\bAi^{y\hi}\partial_{\hi} T\det h_{\hi \hj}]+
%\partial_{\hi}[\tn e^{-2\phi}Vaf' (\bAi)^{\hi y}\partial_y T\det h_{\hi\hj}]+
%\nonumber \\
%+\partial_{\hi}[\tn e^{-2\phi}Vaf' \bAi^{\hi\hj}\partial_{\hj}T\det h_{\hi\hj}]=
=\frac{\tau_{p-1}}{\tau_{non,p}}\partial_y[\tn e^{-2\phi} Vaf'^2(\frac{1}{a^2f'^2}+\partial_{\hat{m}}th^{\hat{m}
\hat{n}}\partial_{\hat{n}}t)\det h_{\hi\hj}]-\nonumber \\
&&-\frac{\tau_{p-1}}{\tau_{non,p}}\partial_y
[\tn e^{-2\phi}V af'\partial_{\hat{m}}
t h^{\hat{m}\hi}\partial_{\hi}t\det h_{\hi\hj}]+\nonumber \\
& &+\frac{\tau_{p-1}}{\tau_{non,p}}\partial_{\hi}[\tn e^{-2\phi}V af'\partial_{\hat{m}}t
h^{\hat{m}\hi}\det h_{\hi\hj}]-\frac{\tau_{p-1}}{\tau_{non,p}}\partial_{\hi}[\tn e^{-2\phi}Vaf'
h^{\hi \hj}\partial_{\hj}t\det h_{\hi\hj}] \ . 
\nonumber \\
\end{eqnarray}
We also used the fact that for (\ref{fans}),(\ref{ansxA}) and (\ref{phihiafT}) 
$\partial_i[N\pi^i\partial_jT\pi^j]$ is equal zero.
%\begin{eqnarray}
%& &\partial_i[N\pi^i\partial_j T\pi^j]
%=\partial_y[N\pi^y \partial_y  T \pi^y]+
%\partial_y[N\pi^y\partial_{\hi}T\pi^{\hi}]+
%\nonumber \\
%&&+\partial_{\hi}[N\pi^{\hi}\partial_y T\pi^y]+
%\partial_{\hi}[N\pi^{\hi}\partial_{\hj}T\pi^{\hj}]=0 \ .
%\nonumber \\
%%\partial_y[(af')^2V]\tn\tpi^{\hi}\partial_{\hi}t
%%\partial_{\hj}t\tpi^{\hj}-\partial_y[(af')^2V]\tn\tpi^{\hi}
%%\partial_{\hi}t\partial_{\hj}t\tpi^{\hj}+\nonumber \\
%%+\partial_{\hi}[\tn (af')^2V\partial_{\hj}t\tpi^{\hj}]
%%-\partial_{\hi}[\tn (af')^2V\partial_{\hj}t\tpi^{\hj}]=0
%%\nonumber \\
%\end{eqnarray}
Since $p_T$ is zero we finally find that the equations of motion for $T$ is
obeyed for the configuration of fields defined above. Note that $t(\xi^{\hi})$ is completely free which is manifestation of spatial diffeomorphism invariance 
of the Hamiltonian formalism.
%We see that in the limit $a\rightarrow \infty$ all terms proportional to 
%$V\frac{dV}{dT}$ goes to zero. The situation is slightly different with the third and fourth where now we have
%\begin{eqnarray}
%+2\tau_p^2V^2 \partial_{\hi}[e^{_2\phi}N]\bAi^{\hi j}\partial_j T
%\det \bA+
%2\tau_p^2e^{-2\phi}NV^2\partial_i[\bAi^{ij}\partial_j T\det \bA]=\nonumber \\
%=2a^2f'^2\tau_p^2V^2\partial_{\hi}[e^{-2\phi}N]
%(-h^{\hi\hj}\partial_{\hj}t+h^{\hi\hat{m}}
%\partial_{\hat{m}}t
%	)\det h_{\hi\hj}+\nonumber \\
%+2\tau_p^2  e^{-2\phi}V^2
%\partial_{\hi}[a^2f'^2(-h^{\hi\hj}\partial_{\hj}t+h^{\hi\hat{m}}\partial_{\hat{m}}t)\det h_{\hi\hj}] +\nonumber \\
%+2\tau_p^2 e^{-2\phi}V^2\partial_y[a^2f'^2(
%-\partial_{\hat{m}}t h^{\hat{m}\hj}\partial_{\hj}t+
%(\frac{1}{a^2f'^2}+\partial_{\hat{m}}th^{\hat{m}\hat{n}}\partial_{\hat{n}}t))\det h_{\hi\hj}]=0\nonumber \\
%\end{eqnarray}
%that vanish as a consequence of the spatial invariance of the world-volume
%section that implies that the position of the kink should not depend on the function $t(\xi)$. 
Let us further  discuss equations of motion for $x^\mu$ that for the 
ansatz (\ref{fans}),(\ref{ansxA}),(\ref{pmuans}) and (\ref{Npres}) reduces into
\begin{eqnarray}
%\partial_0 x^\mu=2NV^\mu_{ \ I}V^\nu_{ \ J}\eta^{IJ}p_\nu+N^{\hi}\partial_{\hi}x^\mu \Rightarrow \nonumber \\
af'V\left(\partial_0 \tx^\mu-2\tn v^\mu_{ \ I}v^\nu_{ \ J}
\eta^{IJ}\tp_\nu-\tn^{\hi}\partial_{\hi}\tx^\nu \right)=0 \nonumber \\
\end{eqnarray}
and we see that this equation is obeyed at the core of the kink $y=t(\xi^{\hi})$  when the expression in the bracket vanishes  which is nothing else then the equation of motion (\ref{eqxp}).

Further, inserting ansatz given above to the equations of motion for $A_{\hi}$ 
% has the form
%\begin{equation}
%\partial_0 A_{\hi}=2Nh_{\hi\hj}\pi^{\hj}-F_{\hi\hk}N^{\hk}-\partial_{\hi}A_{y}N^y+\partial_{\hi}A_0
%\end{equation}
we obtain
\begin{equation}
af'V \left(\partial_0 \ta_i-2\tn h_{\hi\hj}\tpi^{\hj}+F_{\hi\hk}\tn^{\hk}-
\partial_{\hi}\ta_0\right)=0
\end{equation}
that is again obeyed at the core of the kink when the expression in the
bracket that corresponds to (\ref{eqap}), is zero. In case of the equations of motion for $A_y$ we get that it is identically obeyed for $A_y=0$.

%Further, equation of motoin for $A_y$ has the form
%\begin{equation}
%\partial_0 A_y=-F_{y\hi}N^{\hi}=\partial_{\hi}A_y N^y
%\end{equation}
%that we will solve with the ansatz $A_y=0$. Then the equations of motion for $A_{\hi}$ has the same form as equation of motion on D(p-1)-brane. This is also true in case of equation of motion for $x^\mu$. 
Let us now proceed to the equation of motion for $p_\mu$. Again, inserting
ansatz (\ref{fans}),(\ref{ansxA}),(\ref{pmuans}) together with 
(\ref{Nysol}) and using (\ref{phihiafT}) we find that it takes the form
%\begin{eqnarray}
%\partial_0 p_\mu=\pb{p_\mu,H}=
%-Np_\rho \partial_\mu (v^\rho_{ \ I}v^\nu_{ \ J}\eta^{IJ})
%p_\nu \nonumber \\
%+2\partial_{i}[N \pi^{i}h_{\mu\nu}\partial_{\hj} x^\nu \pi^{\hj}]
%-N\pi^{\hi}\partial_{\hi} x^\rho\partial_\mu h_{\rho\sigma}
%\partial_{\hj} x^\sigma -\tau_p^2N\partial_\mu[e^{-2\phi}]V^2
%a^2f'^2\det(h_{\hi\hj})+\nonumber \\
%+2\tau_p^2V^2 a^2f'^2\partial_{\hi}[N e^{-2\phi}h_{\mu\nu}
%\partial_{\hj}x^\nu h^{\hj\hi}\det h_{\hi\hj}]
%-
%\tau_p^2 e^{-2\phi}NV^2a^2f'^2\partial_{\hi} x^\rho \partial_\mu h_{\rho\sigma}
%\partial_{\hj} x^\sigma h^{\hj \hi}\det h_{\hi\hj}-\nonumber \\
%-\partial_i[N^i p_\mu] \Rightarrow  \nonumber \\
%\nonumber \\
%af'V[\partial_0\tp_\mu=
%-\tn p_\rho\partial_\mu(v^\rho_{\ I}v^\nu_{ \ J}\eta^{IJ})\tp_\nu+\nonumber \\
%+2\partial_i[\tn ]
%\end{eqnarray}
\begin{eqnarray}
%\partial_0 p_\mu=\pb{p_\mu,H}=
%-Np_\rho \partial_\mu (v^\mu_{ \ I}v^\nu_{ \ J}\eta^{IJ})
%p_\nu \nonumber \\
%+2\partial_i[N \pi^ih_{\mu\nu}\partial_j x^\nu \pi^j]
%-N\pi^i\partial_i x^\rho\partial_\mu h_{\rho\sigma}
%\partial_j x^\sigma\pi^j -\tau_p^2\partial_\mu[e^{-2\phi}]V^2
%\det(h_{ij}+\partial_i T\partial_j T)+\nonumber \\
%+2\tau_p^2 \partial_i[N e^{-2\phi}V^2 h_{\mu\nu}\partial_j x^\nu \bAi^{ji}
%\det \bA_{ij}]-
%\tau_p^2 e^{-2\phi}NV^2\partial_i x^\rho \partial_\mu h_{\rho\sigma}
%\partial_j x^\sigma \bAi^{ji}\det \bA-\nonumber \\
%-\partial_i[N^i p_\mu] \Rightarrow \nonumber \\
& & \frac{\tau_{non,p}}{\tau_{p-1}}af'V\left(-
\partial_0\tp_\mu
-\tn 
\tp_\rho \partial_\mu (v^\rho_{ \ I}v^\nu_{ \ J}
\eta^{IJ})\tp_\nu+\partial_{\hi}[\tn \tpi^{\hi}h_{\mu\nu}\partial_{\hj}x^\nu\tpi^{\hj}] \right.\nonumber \\
& &
-\tn\tpi^{\hi}\partial_{\hi}\tx^\rho \partial_\mu h_{\rho\sigma}
\partial_{\hj}\tx^\sigma \tpi^{\hj}+2\tau_{p-1}^2\partial_{\hi}[\tn e^{-2\phi}h_{\mu\nu}
\partial_{\hj}\tx^\nu h^{\hj\hi}\det h_{\hi\hj}]-\nonumber \\
& &\left.-\tau_{p-1}^2 
\tn \partial_\mu[e^{-2\phi}]\det h_{\hi\hj}-\tau_{p-1}^2
\tn e^{-2\phi}\partial_{\hi}\tx^\rho \partial_\mu h_{\rho\sigma}
\partial_{\hj}x^\sigma h^{\hj\hi}\det h_{\hi\hj}\right)=0 \ ,  \nonumber \\
\end{eqnarray}
where we used the fact that
\begin{eqnarray}
& &\tau_{non,p}^2\partial_i[Ne^{-2\phi}V^2 h_{\mu\nu}\partial_j x^\nu \bAi^{ji}\det \bA]=
%2\tau_p^2\partial_{\hi}[\tn e^{-2\phi}Vaf' h_{\mu\nu}\partial_{\hj}
%x^\nu h^{\hj\hi}\det h_{\hi\hj}]+\nonumber \\
%+2\tau_p^2 \partial_y[N e^{-2\phi}V a f'h_{\mu\nu}
%\partial_{\hj}x^\nu h^{\hj\hk}\partial_{\hk}t\det h_{\hi\hj}]=
%\nonumber \\
%-2\tau_p^2\partial_y[V a f']\tn e^{-2\phi}h_{\mu\nu}\partial_{\hj}\tx^\nu
%h^{\hj\hi}\partial_{\hi}t\det h_{\hi\hj}
%+2\tau_p^2 \partial_y[V a f']\tn e^{-2\phi}h_{\mu\nu}
%\partial_{\hj}\tx^\nu h^{\hj\hk}\partial_{\hk}t\det h_{\hi\hj}]+
%\nonumber \\
%+2\tau_p^2V af'\partial_{\hi}[N\tn e^{-2\phi}h_{\mu\nu}
%\partial_{\hj}\tx^\nu h^{\hj\hi}\det h_{\hi\hj}] =
%\nonumber \\
2\tau_{non,p}\tau_{p-1}V af'\partial_{\hi}[\tn e^{-2\phi}h_{\mu\nu}
\partial_{\hj}\tx^\nu h^{\hj\hi}\det h_{\hi\hj}] \ ,  \nonumber \\
& &\partial_i[N\pi^i h_{\mu\nu}\partial_{\hj}x^\nu \pi^{\hj}]=\frac{\tau_{non,p}}{\tau_{p-1}}
af'V\partial_{\hi}[\tn \tpi^{\hi}h_{\mu\nu}\partial_{\hj}x^\nu\tpi^{\hj}] \ . 
\nonumber \\
\end{eqnarray}
%and also
%\begin{equation}
%\partial_{\hi}[V^2 a^2 f'^2]=-\frac{d}{dy}[V^2 a^2 f'^2]\partial_{\hi}t
%\end{equation}
%together with
%\begin{eqnarray}
%\partial_i[N\pi^i h_{\mu\nu}\partial_{\hj}x^\nu \pi^{\hj}]=\nonumber \\
%%=\partial_y[\tn (af'V)\partial_{\hk}t\tpi^{\hk}h_{\mu\nu}
%%\partial_{\hj}x^\nu \tpi^{\hj}]+
%%\partial_{\hi}[\tn (af'V)\tpi^{\hi}h_{\mu\nu}\partial_{\hj}x^\nu
%%\tpi^{\hj}]=\nonumber \\
%%=\partial_y[af'V]\tn \partial_{\hk}t\tpi^{\hk}h_{\mu\nu}\partial_{\hj}
%%x^\nu\tpi^{\hj}-\partial_y[af'V]\partial_{\hi}t 
%%\tn \tpi^{\hi}h_{\mu\nu}\partial_{\hj}x^\nu\tpi^{\hj}+
%%af'V\partial_{\hi}[\tn \tpi^{\hi}h_{\mu\nu}\partial_{\hj}x^\nu\tpi^{\hj}]=
%%\nonumber \\
%=af'V\partial_{\hi}[\tn \tpi^{\hi}h_{\mu\nu}\partial_{\hj}x^\nu\tpi^{\hj}]
%\nonumber \\
%\end{eqnarray}
In summary, we showed that fields given in (\ref{fans}),(\ref{ansxA}) and (\ref{pmuans}) solve equations of motion for non-BPS Dp-brane in $k-$brane
Carrollian background on conditions when the 
 fluctuations around kink solution  obey  canonical equations of motion for D(p-1)-brane in the same background.  
\subsection{Fundamental String Solutions}
Let us now discuss solutions when the tachyon field on the world-volume of unstable Dp-brane is in its vacuum state corresponding to $V(T_{min})=
\frac{dV}{dT}(T_{min})=0$. This situation was  extensively studied 
in case of relativistic tachyon effective action  
\cite{Sen:2003bc,Kim:2003he,Gibbons:2000hf,Kluson:2017fam,Kluson:2016tgp,Kwon:2003qn} and in case of the  flat Carroll Dp-bran in \cite{Kluson:2017fam}. Our analysis
will closely follow \cite{Sen:2003bc} and also \cite{Kluson:2017fam}. 

To begin with we again write equations of motion for Dp-brane in $k-$brane Carrollian background at the tachyon vacuum state. In this case the equations of motion simplify considerably
\begin{eqnarray}\label{eqvac}
& &\partial_0 x^\mu=
2NV^\mu_{ \ I}V^\nu_{ \ J}\eta^{IJ}p_\nu+N^i\partial_i x^\mu \ , 
\nonumber \\
& &\partial_0 p_\mu=
-Np_\rho \partial_\mu (v^\rho_{ \ I}v^\nu_{ \ J}\eta^{IJ})
p_\nu \nonumber \\
& &+2\partial_i[N \pi^ih_{\mu\nu}\partial_j x^\nu \pi^j]
-N\pi^i\partial_i x^\rho\partial_\mu h_{\rho\sigma}
\partial_j x^\sigma-\partial_i[N^i p_\mu] \ , \nonumber \\
& &\partial_0 \pi^i=\partial_k[N^k\pi^i]-\partial_l[N^i\pi^l] \ , 
\nonumber \\
& &\partial_0 A_i=
2Nh_{ij}\pi^j-F_{ij}N^j+\partial_i A_0 \ , \nonumber \\
& &p_\mu v^\mu_{ \ I}v^\nu_{ \ J}\eta^{IJ}p_\nu+\pi^ih_{ij}\pi^j=0 \ , \quad 
p_\mu\partial_i x^\mu+F_{ij}\pi^j=0 \ . \nonumber \\
\end{eqnarray}
We would like to show that these equations of motion can be interpreted as
equations of motion for fundamental string in $k-$brane Carrollian background. To begin with let us consider string action in the canonical form
\begin{equation}
S=\int d\tau d\sigma (K_\mu \partial_\tau Z^\mu-\tlambda_\tau \mK_\tau-
\tlambda_\sigma \mK_\sigma) \ , 
\end{equation}
where
\begin{equation}
\mK_\tau=K_\mu g^{\mu\nu}K_\nu+\ttau_F^2 \partial_\sigma Z^\mu
\partial_\sigma Z^\nu g_{\mu\nu} \ , \quad 
\mK_\sigma=K_\mu \partial_\sigma Z^\mu \ , 
\end{equation}
and where $K_\mu(\tau,\sigma)$ are momenta conjugate to $Z^\mu(\tau,\sigma)$
and where the world-sheet is labeled by $\tau$ and $\sigma$. Let us consider
 Carroll expansion of the metric
\begin{equation}
g_{\mu\nu}=c^2T_\mu^{ \ \tilde{I}}T_\nu^{\ \tilde{J}}\eta_{\tilde{I}\tilde{J}}+h_{\mu\nu} \ , \quad 
g^{\mu\nu}=\frac{1}{c^2}V^\mu_{ \ \tilde{I}}V^\nu_{ \ \tilde{J}}\eta^{\tilde{I}\tilde{J}}+\Pi^{\mu\nu} \ , 
\end{equation}
where $\tilde{I},\tilde{J}$ vary from $0,\dots,l$ where generally $l$ is different from $k$. Then we also rescale $\tN$ and $\ttau_F$ as
\begin{equation}
\tlambda_\tau=c^2\lambda_\tau \ , \quad \tlambda_\sigma=\lambda_\sigma \ , \quad \ttau_F=\frac{1}{c}\tau_F \ . 
\end{equation}
As a result we obtain an action for string in $l-$brane Carrollian background in the form 
\begin{eqnarray}
S=\int d\tau d\sigma
(K_\mu\partial_0 Z^\mu-\lambda_\tau(K_\mu v^\mu_{ \ \tilde{I}}v^\nu_{ \ \tilde{J}}\eta^{\tilde{I}\tilde{J}}K_\nu
+\tau_F^2h_{\sigma\sigma})-\lambda_\sigma K_\mu \partial_\sigma Z^\mu) \ . \nonumber \\
\end{eqnarray}
From this action we derive corresponding equations of motion 
\begin{eqnarray}
& &\partial_\tau Z^\mu=2\lambda_\tau v^\mu_{\ \tilde{I}}v^\nu_{ \ \tilde{J}}\eta^{\tilde{I}\tilde{J}}K_\nu+\lambda_\sigma \partial_\sigma Z^\mu
\ , \nonumber \\
& &-\partial_\tau K_\mu=-\lambda_\tau K_\rho \partial_\mu (v^\rho_{ \ \tilde{I}}
\eta^{\tilde{I}\tilde{J}})K_\nu
-\tau_F^2\lambda_\tau \partial_\sigma Z^\rho \partial_\mu h_{\rho\sigma}
\partial_\sigma Z^\sigma+\nonumber \\
& &+2\tau_F^2\partial_\sigma[\lambda_\tau \partial_\sigma Z^\nu
h_{\nu\mu}]+\partial_\sigma[\lambda_\sigma K_\mu] \ , \nonumber \\
& &K_\mu v^\mu_{ \ \tilde{I}}v^\nu_{ \ \tilde{J}}\eta^{\tilde{I}\tilde{J}}K_\nu
+\tau_F^2h_{\sigma\sigma}=0 \ , \quad 
 K_\mu \partial_\sigma Z^\mu=0 \ .
\nonumber \\
\end{eqnarray}
In order to interpret the equations of motion for non-BPS Dp-brane at the tachyon
vacuum (\ref{eqvac}) as the equations of motion for Carrollian string  we immediately see that the range of $I$ should coincide with $\tilde{I}$ so that $k=l$. Then we
proceed in similar way as in \cite{Kluson:2017fam}. First of all we introduce
projector
\begin{equation}
\triangle^i_{ \ j}=
\delta^i_j-\frac{\pi^i h_{jk}\pi^k}{\pi^m h_{mn}\pi^n}
\end{equation}
that obeys
\begin{equation}
\triangle^i_{ \ j}\pi^j=0 \ , \quad 
\triangle^i_{ \ j}\triangle^j_{ \ k}=\triangle^i_{ \ k} \ . 
\end{equation}
Then we split $N^i$ as
\begin{equation}
N^i=N^j\delta_j^i=N^j(\triangle_j^i+
\frac{\pi^i h_{jk}\pi^k}{\pi^m h_{mn}\pi^n})=N^i_{\bot}+\pi^i N_{II} \ , 
\end{equation}
where by definition 
\begin{equation}
\pi^i h_{ij}N^j_{\bot}=0 \ . 
\end{equation}
As the next step we introduce dimensionless vector $\omega^i$ through the relation
\begin{equation}
\pi^i=\tau_{non,p}\omega^i \ . 
\end{equation}
and search for solution when $\pi^i$ is constant. In fact, this ansatz clearly solves equation $\partial_i \pi^i=0$ while the equations of motion for $\pi^i$ take the form
\begin{eqnarray}\label{eqpi}
%\partial_0 \pi^i=\pb{\pi^i,H}=\partial_k[N^k\pi^i]-\partial_l[N^i\pi^l] \Rightarrow
%\nonumber \\
%0=\partial_k[N^k]\pi^i-\partial_l[N^i]\pi^l=
%\partial_k N^k_{\bot}\pi^i+\pi^k\partial_k N_{II}\pi^i
%-\partial_l[N^i_{\bot}]\pi^l-\pi^l\partial_l N_{II}\pi^i
%=\nonumber \\
0=\partial_0\pi^i=\partial_k N^k_{\bot}\pi^i-\partial_\sigma N^i_{\bot}\tau_{non,p} \ ,  \nonumber \\
\end{eqnarray}
where we introduced $\sigma$ coordinate through the relation 
\begin{equation}
\partial_\sigma\equiv \omega^i\partial_i \ . 
\end{equation}
We solve the equation (\ref{eqpi}) above by imposing  $N^k=\mathrm{const}$. Note however that $N_{II}$ is still non-specified so that there is still spatial dimensional invariance along $\sigma$ direction.
 We further presume that all world-volume fields depend on $\sigma$ only. Then when we perform identification
\begin{equation}
\pi_\mu(\xi^0,\sigma)|_{\xi^0=\tau}=\frac{\tau_{non,p}}{\tau_F}K_\mu(\tau,\sigma) \ , 
\quad x^\mu(\xi^0,\sigma)|_{\xi^0=\tau}=Z^\mu(\tau,\sigma)
\end{equation}
we see that the equations of motion for non-BPS Dp-brane at the tachyon vacuum reduces to the equations of motion for string in $k-$brane Carrollian background. Clearly the dynamics of this string is again trivial however this result shows consistency of the tachyon vacuum condensation in case of non-BPS Dp-brane in $k-$brane Carrollian background. 

{\bf Acknowledgement:}
\\
This work 
is supported by the grant “Integrable Deformations”
(GA20-04800S) from the Czech Science Foundation
(GACR).

\end{document}